\documentclass{article}

\usepackage{arxiv}

\usepackage[utf8]{inputenc} 
\usepackage[T1]{fontenc}    
\usepackage{hyperref}       
\usepackage{url}            
\usepackage{booktabs}       
\usepackage{amsfonts}       
\usepackage{nicefrac}       
\usepackage{microtype}      
\usepackage{lipsum}
\usepackage{graphicx}
\graphicspath{ {./images/} }

\usepackage{lmodern}
\usepackage{amssymb,amsmath}
\usepackage{ifxetex,ifluatex}

\usepackage{xcolor}

\usepackage{longtable,booktabs,array}
\usepackage{calc} 
\usepackage{etoolbox}
\makeatletter
\patchcmd\longtable{\par}{\if@noskipsec\mbox{}\fi\par}{}{}
\makeatother
\IfFileExists{footnotehyper.sty}{\usepackage{footnotehyper}}{\usepackage{footnote}}
\makesavenoteenv{longtable}
\makeatletter
\def\maxwidth{\ifdim\Gin@nat@width>\linewidth\linewidth\else\Gin@nat@width\fi}
\def\maxheight{\ifdim\Gin@nat@height>\textheight\textheight\else\Gin@nat@height\fi}
\makeatother
\setkeys{Gin}{width=\maxwidth,height=\maxheight,keepaspectratio}
\makeatletter
\def\fps@figure{htbp}
\makeatother
\title{Transformer-CNN Fused Architecture for Enhanced Skin Lesion Segmentation}

\author{
 Siddharth Tiwari \\
  School of Computer Science and Mathematics\\
  Liverpool John Moores University\\
}

\begin{document}
\maketitle
\begin{abstract}
The segmentation of medical images is important for the improvement and creation of healthcare systems, particularly for early disease detection and treatment planning. In recent years, the use of convolutional neural networks (CNNs) and other state-of-the-art methods has greatly advanced medical image segmentation. However, CNNs have been found to struggle with learning long-range dependencies and capturing global context due to the limitations of convolution operations. In this paper, we explore the use of transformers and CNNs for medical image segmentation and propose a hybrid architecture that combines the ability of transformers to capture global dependencies with the ability of CNNs to capture low-level spatial details. We compare various architectures and configurations and conduct multiple experiments to evaluate their effectiveness. 
\end{abstract}


\section{Introduction}
\label{sec:Introduction}
There has been a significant effort in the literature to improve skin lesion segmentation methods. Traditional methods that rely on hand-crafted features have been found to be inflexible and often lead to poor segmentation performance for lesions with various variations. To address this issue, convolutional neural network (CNN) models have been proposed and have shown promising results compared to traditional methods. However, CNN-based models are still not sufficient to fully address the challenges of skin lesion segmentation due to their lack of global context. Recently, using transformers and self-attention mechanisms to consider global context while segmenting images has been suggested. While transformers excel at global context, they struggle with fine-grained details, particularly in medical imaging. To address this, efforts have been made to combine CNNs and transformers to take advantage of both methods and model both low-level features and global feature interaction.

\subsection{Traditional Approaches to Skin Lesion Segmentation}
Conventional techniques for skin lesion segmentation can be divided into several categories, including thresholding methods which use threshold values to distinguish between healthy and affected tissue \cite{al2018skin}, clustering methods which use color-space features for clustering regions \cite{Dhanachandra2015}, region-based techniques that focus on region analysis \cite{Wong2011}, active contour techniques that use evolutionary algorithms for segmentation \cite{Kasmi2016}, and supervised learning methods such as SVM and ANNs \cite{Li2018}. Despite these efforts, skin lesion segmentation remains a difficult task due to the complex and varied features of these images. Table \ref{tab:tab1} outlines some of the characteristics that contribute to the challenges faced by traditional methods.

\begin{center}
\begin{longtable}[]{@{}
  >{\centering\arraybackslash}p{(\columnwidth - 4\tabcolsep) * \real{0.33}}
  >{\centering\arraybackslash}p{(\columnwidth - 4\tabcolsep) * \real{0.33}}
  >{\centering\arraybackslash}p{(\columnwidth - 4\tabcolsep) * \real{0.33}}@{}}

\includegraphics[width=1.94792in,height=1.45833in]{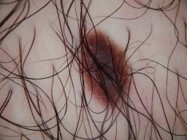}

(a)Hairs &
\includegraphics[width=1.9375in,height=1.45833in]{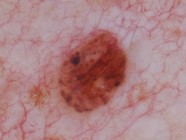}

b) Blood Vessels &
\includegraphics[width=1.9375in,height=1.45833in]{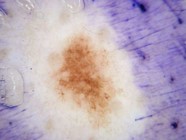}

(c) Surgical masking \\

\endhead
\includegraphics[width=1.9375in,height=1.51389in]{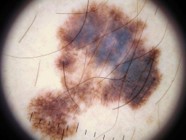}

(d) Irregular border and black frame &
\includegraphics[width=1.9375in,height=1.51389in]{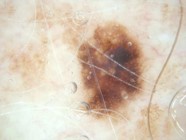}

(e) Bubbles &
\includegraphics[width=1.95139in,height=1.51389in]{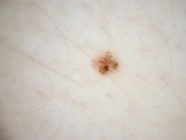}

(f) Very Small Lesion \\
\includegraphics[width=1.91667in,height=1.49653in]{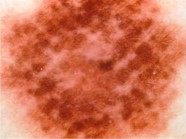}

(g) Very Large Lesion &
\includegraphics[width=1.91667in,height=1.49653in]{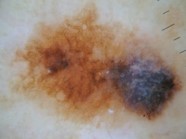}

(h) Fuzzy border and variegated colouring &
\includegraphics[width=1.90972in,height=1.5in]{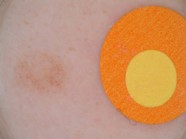}

(i) Low Contrast and Colour calibration Chart \\

\caption{Factors complicating segmentation of dermoscopy images.}
\label{tab:tab1}
\end{longtable}
\end{center}

\subsection{Deep Learning for Skin Lesion Segmentation}
Deep learning has greatly improved the performance of medical image tasks, including skin lesion segmentation. These methods have the ability to extract deep features from complex datasets and have been successful in medical image segmentation tasks using deep convolutional networks. These networks can learn to detect and segment fine-grained features and extract hierarchical features related to the appearance and semantics of images in large datasets.

\subsubsection{CNNs for Skin Lesion Segmentation}
CNNs have achieved exceptional results in numerous medical image segmentation tasks by training end-to-end models for task-specific hierarchical feature representation. While they have been successful, they struggle to efficiently capture global context information. Some previous works have attempted to include global information by using large receptive fields and downsampling, resulting in deep networks. However, this approach has several drawbacks: 1) training very deep networks can result in diminishing feature reuse, 2) reducing spatial resolution can discard local information critical for dense prediction tasks, 3) training parameter-heavy deep networks with small image datasets can lead to overfitting and unstable networks.

\subsubsection{Transformers for Skin Lesion Segmentation}
There has been a growing interest in using transformer-based architectures for computer vision tasks, particularly after their success in natural language processing. One of the earliest models in this area was the Vision Transformer (ViT), which achieved results comparable to CNN-based architectures by inputting images as a sequence of patches. Since the ViT, several other models have been proposed, including TNT \cite{Xu2022}, Swin \cite{liu2021swin}, SwinV2 \cite{liu2022swin}, XCiT \cite{NEURIPS2021_a655fbe4}, CaiT \cite{touvron2021going}, BeiT \cite{bao2021beit}, DeiT \cite{touvron2021training}, iBOT \cite{zhou2021ibot}, and DINO \cite{caron2021emerging}. These models have shown promising results and have benefited from advances in training methods. ViTs trained on ImageNet demonstrate higher shape distortion compared to CNN models of similar capacity \cite{naseer2021intriguing} and are capable of achieving human-level shape-warping performance when trained on a stylized version of ImageNet (SIN). Some models can even use conflicting cues, such as shape and texture warping, with different tokens.

\subsubsection{Combining CNNs and Transformers for Skin Lesion Segmentation}
Past efforts to integrate CNNs and transformers have primarily focused on using stacked transformer layers or vanilla transformer layers in a sequential manner to replace CNNs. For example, the SEgmentation TRansformer (SETR) \cite{zheng2021rethinking} uses transformer encoders and demonstrates state-of-the-art performance in segmentation tasks. TransUnet \cite{chen2021transunet} uses CNNs to extract low-level features, which are then input to transformers to learn global feature interaction. Transfuse \cite{zhang2021transfuse} is based on the vision transformer architecture ViT and uses a fusion module to combine features extracted by CNNs and transformers, while MedT \cite{valanarasu2021medical} relies on an axial-attention transformer and explores the feasibility of using transformers even with limited large-scale datasets. These models show the potential of transformers for medical imaging segmentation, but they mostly use transformers as encoders, and the effectiveness of transformers as decoders has yet to be fully explored.

Another promising approach for image segmentation is the use of multi-scale feature representations. For example, Cross-Attention Multi-Scale Vision Transformer (CrossViT) \cite{chen2021crossvit} extracts multi-scale features using a dual-branch transformer, and Multi Vision Transformers (MViT) \cite{fan2021multiscale} use multi-scale feature hierarchies with transformer models for both images and videos. Multi-modal Multi-scale TRansformer (M2TR) \cite{wang2022m2tr} detects local inconsistency at different scales through a proposed multi-scale transformer. Overall, multi-scale feature representations have not been widely used in image segmentation, despite their effectiveness for vision transformers.

\section{Proposed Method}
\label{sec:Proposed Method}

\paragraph{Modeling}
To address the limitations of CNNs in capturing global context information, we propose a novel architecture that combines a CNN encoder and a parallel transformer-based segmentation network. The combined architecture allows for the learning of low-level spatial features from the CNN and high-level semantic context from the transformer. This approach avoids the need for very deep architectures, which can suffer from gradient vanishing, and results in a smaller model with the same learning capacity and optimized inference speed, suitable for deployment on low compute edge devices.

The proposed model has a dual branch parallel architecture that processes information in two different ways. The CNN branch gradually increases the receptive field, while the transformer branch starts with a global self-attention mechanism to recover local details at the end. A proposed fusion module selectively combines the extracted features from both branches with the exact resolution. A gated skip-connection is used to merge multi-level feature maps and generate the segmentation masks.

The transformer branch follows a general encoder-decoder architecture. The input image is divided into evenly sized patches, which are flattened and passed through a linear embedding layer. The image shape of \(H \times W \times 3\) can be divided into patches using the parameter (S) such that the output resolution of the patches is \(N = H/S \times W/S\). In vision transformers, the value of S is typically set to 16. The patches are flattened and input to a linear embedding with output dimension \(D_{o}\)), resulting in a raw embedding sequence \(e \in \mathbb{R}^{N \times D_{0}}\). A positional embedding with learnable parameters of the same dimension is added to incorporate spatial prior information. This embedding is fed through a transformer encoder with (L) layers of MSA (Multi-Headed Self-Attention) and MLP (Multi-layer Perceptron). The self-attention mechanism aggregates global information in each layer while simultaneously updating the states of the embedded patches.

The attention mechanism is based on a trainable memory with key and value vector pairs. A set of k key vectors (packed together into a \(matrix\ K \in R^{k \times d}\))) are compared to a query vector  \(q \in \ R^{d}\) using inner products. These inner products are then normalized and scaled using a SoftMax function to produce k weights. The weighted sum of a group of k value vectors (packed into \(V\  \in \ R^{k \times d}\)) is the result of the attention. It produces an output matrix  (of dimension \(N\  \times \ d\)) for a sequence of N query vectors packed into \(Q\  \in \ R^{k \times d}\):

\begin{center}
$Attention(Q,K,V) = Softmax\left(\frac{QK^{T}}{\sqrt{d}}\right)V$
\end{center}

where the  \(\sqrt{}d\) term normalizes the input and the SoftMax function is applied to each row of the matrix.

In the case of self-attention, the query, key, and value matrices are independently computed from a set of N input vectors  (packed into
\(X\  \in \ R^{N \times D}\)) as follows: \(Q\  = \ XW_{Q}\),
\(K\  = \ XW_{K},\) \(V\ \  = \ XW_{V}\), using linear transformations \(W_{Q}\), \(W_{K}\), \(W_{V}\) with the constraint \(k\  = \ N\), indicating that the attention is applied between all input vectors.

The Multi-head Self-Attention (MSA) layer is created by considering h attention "heads" or  \(h\) self-attention functions applied to the input. Each head produces a size \(N\  \times \ d\) sequence, which are rearranged into a \(N\  \times \ dh\)  sequence and then projected through a linear layer into a \(N\  \times \ D\) sequence. The output of the transformer layer, the encoded sequence \(\ z^{L}\  \in \ R^{N \times D}\), is then normalized using layer normalization.

For the decoder part of the proposed model, we plan to use the progressive up sampling (PUP) method, as used in the SEgmentation TRansformer (SETR). From a traditional CNN pipeline, we remove the last block and utilize the transformer to capture global context information while benefiting from its advantages.

In the proposed architecture, the CNN branch uses ResNet-34 (R34) as its backbone, while the transformer branch uses DeiT-Small (DeiT-S) as its backbone. This allows for the retention of richer local information while resulting in a shallower model. Since ResNet-based models typically contain five blocks, with feature maps being downsampled by a factor of two in each block, the outputs from the second, third, and fourth blocks are taken and fused with the outputs from the same blocks of the transformer branch. The resulting parallel CNN-transformer architecture is trained in an end-to-end fashion.

\begin{figure}
\begin{center}
\includegraphics[width=6.5in,height=4.86667in]{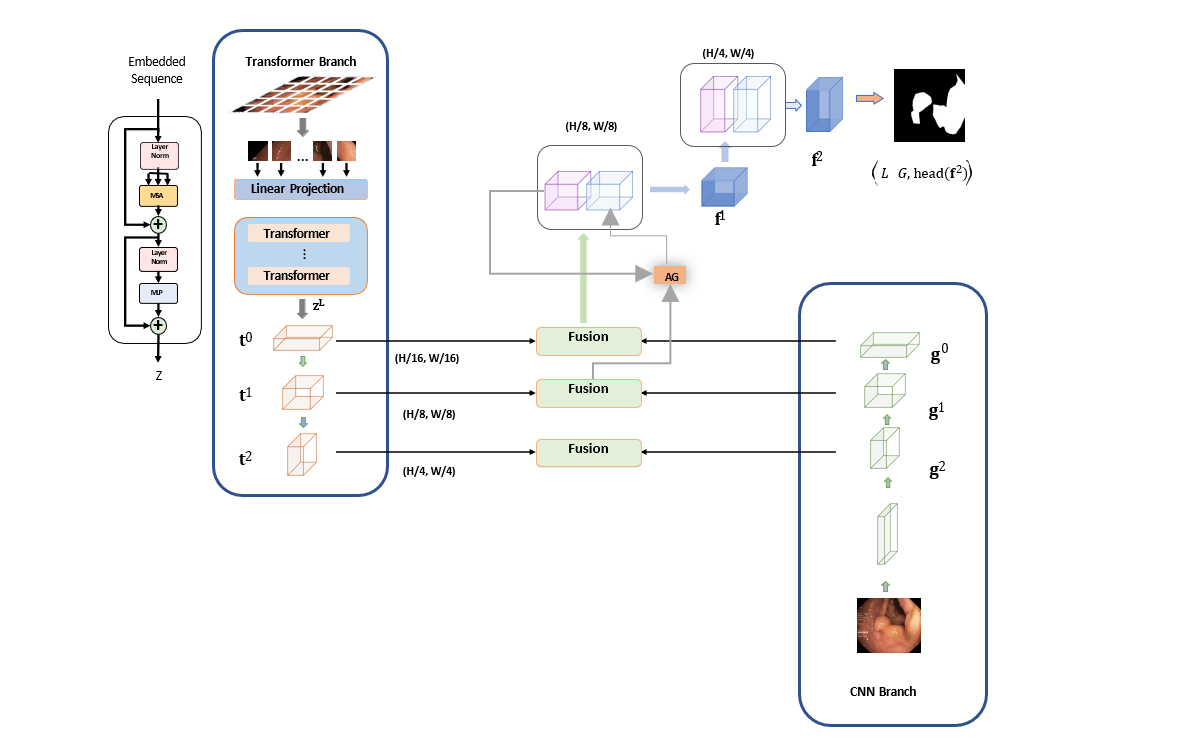}
\caption{Logical flow of the Architecture}
\label{fig:fig1}
\end{center}
\end{figure}

\section{Implementation}
\label{sec:Implementation}

\subsection{Data Selection}
\label{data-selection}
The dataset selected for this research work is the ISIC dataset published and
collected by the International Skin Imaging Collaboration (ISIC). This
dataset contains a collection of dermoscopic images of skin lesions. The
dataset has been collected from major international clinical centres and
from various devices within each centre. All images in the dataset have
been screened for quality assurance and privacy. The Task 1 of the ISIC
2017\cite{codella2018skin} challenge provides a dataset for skin-lesion segmentation with
2,000 images for training with ground truth masks (binary images). For
validation, there are 150 images, and for testing, there are 600 images.
Example images of the dataset are given for reference in Table \ref{tab:samples}

\begin{center}
\begin{longtable}{@{}
  >{\centering\arraybackslash}p{(\columnwidth - 12\tabcolsep) * \real{0.5}}
  >{\centering\arraybackslash}p{(\columnwidth - 12\tabcolsep) * \real{0.5}}@{}}

\includegraphics[width=1.94792in,height=1.45833in]{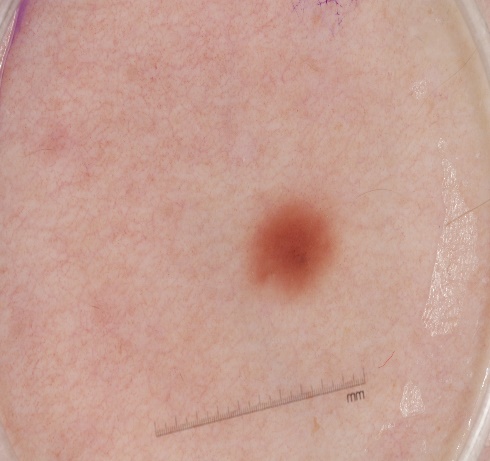} &
\includegraphics[width=1.94792in,height=1.45833in]{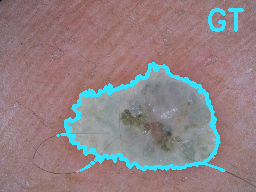} \\
(a) Sample & (b) Sample ground truth \\

\caption{Sample images from the dataset}
\label{tab:samples}
\end{longtable}
\end{center}

\subsection{Data Pre-processing and Transformation}
All images are resized to 192×256 following the setting in \cite{al2018skin}. Different data augmentation strategies are expected to be
used, like random rotation, colour jittering, horizontal flip, etc. and
applied during training.

Pre-processing may facilitate the segmentation of skin lesion images.
These pre-processing operations include:

\begin{itemize}
\item
  \textbf{Downsampling}: While dermoscopy results in a high-resolution
  image with large sizes, many Convolutional Neural Networks (CNNs) take
  fixed-size input images, mostly 224x224 or 299x299 pixels. Even those
  CNNs which can handle variable size images (e.g., fully folded
  networks, FCN) benefit from downsampling as these are computationally
  easier to process. That is why downsampling is a very common operation
  found in the literature for skin lesion segmentation..
\end{itemize}

\begin{itemize}
\item
  \textbf{Colour Space Transformations}: Most models expect RGB images
  from most models, while some works use alternative colour spaces like
  HSV, CIELUV and CIELAB. RGB channels are often combined with some of
  these, including decoupling luminance, increasing class separability,
  chrominance, eliminating highlights and achieving invariance to
  lighting or viewing angle.
\item
  \textbf{Contrast enhancement}: A significant reason for segmentation
  errors is insufficient contrast (Table. 1(i) \ref{tab:tab1}), leading to some work to
  improve image contrast before segmentation.
\item
  \textbf{Colour normalization}: Inconsistencies can also be caused due
  to variation in illumination. Colour normalisation can be used to
  address this issue.
\item
  \textbf{Artifact removal}: Dermoscopic images often show artefacts,
  including hairs, leading some studies to remove them before
  segmentation.
\end{itemize}

\subsection{Architecture Implementation}
\label{Architecture Implementation}
We follow a general encoder-decoder architecture for the transformer
branch. We divide the input image into evenly divided patches which are
flattened for passing into a linear embedding layer. An image of shape
\(H \times W \times 3\) is divided into patches with a parameter \(S\)
such that \(\frac{H}{S} \times \frac{W}{S}\) is the output resolution of
the divided patches. Here, we follow a Vision transformer based
architecture, so the value of \(S\) is set to 16. A positional embedding
with the same dimension is used to take advantage of the spatial prior.
This embedding is fed to a transformer encoder having multiple layers of
MSA (Multi-Headed Self-Attention) and MLP (Multi-layer Perceptron). To
obtain the encoded sequence, the output of the transformer layer is
normalised using Layer normalisation. We use Progressive Upsampling
(PUP) method for the decoder part, as used in SETR \cite{zheng2021rethinking}. From the usual CNNs
pipeline, we remove the last block and take advantage of the Transformer
considering the benefits brought by it to obtain global context
information.

ResNet-34 (R34) is used as backbone the CNN branch and DeiT-Small
(DeiT-S) which has 8 layers can be used as backbone of the transformer
branch. This will help retaining richer local information and the output
will be a shallower model. Since models based on ResNet normally contain
5 blocks and the feature maps are down sampled by a factor of 2 in each
block, outputs from the second, third and fourth blocks is taken to fuse
with the outputs from the same blocks of transformer branch. This
parallel CNN transformer architecture is trained end-to-end.

\subsubsection{Fusion Module Implementation}\label{fusion-module-implementation}
We suggest a new Fusion module (see Figure \ref{fig:fig1}) that combines
self-attention and multi-modal fusion techniques in order to
successfully fuse the encoded information from CNNs and Transformers.
Specifically, we do the following operations to get the fused feature
representation \(f^{i},\ i\  = \ 0,\ 1,\ 2\) :

\begin{align*}
{\hat{t}}^{i} &= ChannelAttn(t^{i}) \\
{\hat{g}}^{i} &= SpatialAttn(g^{i}) \\
{\hat{b}}^{i} &= Conv\left(t^{i}W_{1}^{i}\bigodot g^{i}W_{2}^{i}\right) \\
f^{i} &= Residual\left(\left[{\hat{b}}^{i},{\hat{t}}^{i},{\hat{g}}^{i}\right]\right)
\end{align*}

where \textbar{}\(\bigodot|\)  is the Hadamard product and Conv is a
3\(\times 3\) convolution layer and where \(W_{1}^{i}\) \(\in\)
\(\mathbb{R}^{D_{i} \times L_{i}}\) and \(W_{2}^{i}\) \(\in\)
\(\mathbb{R}^{C_{i} \times L_{i}}\). The SE-Block implementation of the
channel attention promotes global information from the Transformer
branch. Due to the possibility of low-level CNN features becoming noisy,
spatial attention is utilised as spatial filters to highlight local
details and suppress irrelevant regions. The fine-grained interaction
between features from the two branches is then modelled by the Hadamard
product. The interaction features \({\hat{b}}^{i}\ \)and attended
features
\({\hat{t}}^{i},\ {\hat{g}}^{i}\text{\ are\ then\ concatenated\ and\ transmitted\ through}\)
a Residual block. For the current spatial resolution, the resulting
feature \(f^{i}\) successfully captures both the global and local
context.

Finally, \({\hat{f}}^{i}\)'s are combined using the attention-gated (AG)
skip-connection to generate final segmentation,we have:
\begin{center}
${\hat{f}}^{i+1} = Conv\left(\left[Up({\hat{f}}^{i}), AG({\hat{f}}^{i+1}, Up({\hat{f}}^{i}))\right]\right)$ and ${\hat{f}}^{0} = f^{0}$
\end{center}

\subsubsection{Loss Function}\label{loss-function}
Boundary pixels are given more weight when training the entire network
using the weighted IoU loss and binary cross entropy loss
\(L = \ L_{\text{IoU}}^{w}\)+ \(L_{\text{bce}}^{w}\). By immediately
scaling the input feature maps to their original resolution and applying
convolution layers to create M maps, where M is the number of classes, a
simple head can produce segmentation predictions. By further supervising
the transformer branch and the initial fusion branch, we apply deep
supervision to enhance the gradient flow. The final training loss is
given by 

\begin{center}
$\mathcal{L} = \alpha L(G,head({\hat{f}}^{2})) + \gamma L(G,head(t^{2})) + \beta L(G,head(f^{0}))$
\end{center}

where
\(\alpha,\beta\ and\ \gamma\ \)are trainable hyperparameters and \emph{G}
is the ground truth.

\subsubsection{Implementation details}\label{implementation-details}
All the discussed models use weights pre-trained on ImageNet. A single
NVIDIA-RTX-2060 GPU was used for training the model using
using the PyTorch framework.The below-mentioned training strategies are
applied for model training.
All the training images are first normalised using Pytorch normalize
transform with below ImageNet Mean:
\begin{center}
Normalize\(\begin{pmatrix}
\centering
0.485 & 0.456 & 0.406 \\
0.229 & 0.224 & 0.225 \\
\end{pmatrix}\)
\end{center}

Then we apply below transformations from the albumentations library :

\begin{table}[ht]
\caption{Table of data augmentation techniques}
\label{tab:aug}
\centering
\begin{tabular}{ll}
\toprule
Technique & Parameters \\
\midrule
ShiftScaleRotate & shift\_limit=0.15, scale\_limit=0.15, rotate\_limit=25, \\
& p=0.5, border\_mode=0 \\
ColorJitter &  \textbf{-} \\
HorizontalFlip &  \textbf{-} \\ 
VerticalFlip &  \textbf{-}  \\ 
\bottomrule
\end{tabular}
\end{table}

The values of \(\alpha,\beta\ and\ \gamma\ \) are set at 0.5,0.3 and 0.2, respectively. 
The model was trained for 30 epochs with a batch size of 16, using the Adam
optimiser with a learning rate of 1e-4. A small learning rate of 7e-5
was finally used for training the model. The model is evaluated on the
validation dataset during training. The typical evaluation metrics for
the segmentation task are used to conduct a quantitative analysis of the
outcomes.

\subsubsection{Evaluation Metrics}\label{evaluation-metrics}

The problem of segmentation of skin lesions can be framed as a
pixel-by-pixel binary classification task. Here, the negative class
corresponds to the background skin, and the positive class corresponds
to the lesion. Given an input image and automated and manual image
segmentation, we can define quantitative scores using a combination of
true positives, false positives, true negatives, and false negatives.

Suppose we have a pair of manual and automated segmentations, we
construct a 2x2 confusion matrix \(C = (\begin{matrix}
\text{TP} & \text{FN} \\
\text{FP} & \text{TN} \\
\end{matrix}\)), where TP is True positives, FP is False Positives, FN
is False Negatives, and TN is True Negatives. If total number of pixels
in each frame is denoted by N, then we have N = TP + FP + FN + TN. Using
these quantiles, various similarity measures can be defined for the
quantification of the accuracy of the segmentation.

\begin{align*}
Accuracy &= \frac{TP + TN}{TP + FN + FP + TN} \\
Precision &= \frac{TP}{TP + FP} \\
Recall &= \frac{TP}{TP + FN} \\
Sensitivity &= \frac{TP}{TP + FN} \\
Specificity &= \frac{TN}{TN + FN} \\
F-measure &= \frac{2TP}{2TP + FP + FN} \\
Jaccard\ index &= \frac{TP}{TP + FN + FP} \\
Matthews\ Correlation\ Coefficient &= \frac{TP \cdot TN - FP.FN}{\sqrt{(TP + FP)(TP + FN)(TN + FP)(TN + FN)}}
\end{align*}

\section{Experiments and Results}
\subsection{Performance Evaluation and Comparison with
State-of-the-Arts}
On the ISIC 2017 test set, the ISBI 2017 challenge ranked approaches
using the Jaccard Index. As evaluation metrics, we use the Jaccard
Index(IOU), Dice score, and pixel-wise accuracy in this case. The table
below shows the comparative metrics against leading techniques.

\begin{table}[ht]
\caption{Results visualization on 4 selected images with failures}
\label{tab:results}
\centering
\begin{tabular}{cccccc}
\toprule
Methods & Backbones & Epochs & Jaccard & Dice & Accuracy \\
\midrule
CDNN & \textbf{-} & \textbf{-} & 0.765 & 0.849 & 0.934 \\
DDN & ResNet-18 & 600 & 0.765 & 0.866 & 0.939 \\
FrCN & VGG16 & 200 & 0.771 & 0.871 & 0.94 \\
DCL-PSI & ResNet-101 & 150 & 0.777 & 0.857 & 0.941 \\
SLSDeep & ResNet-50 & 100 & 0.782 & 0.878 & 0.936 \\
Unet++ & ResNet-34 & 30 & 0.775 & 0.858 & 0.938 \\
Developed-Model & R34+DeiT-S & \textbf{30} & \textbf{0.795} & \textbf{0.872} & \textbf{0.944} \\
\bottomrule
\end{tabular}
\end{table}

Without any pre- or post-processing, the produced model has a Jaccard
score that is 1.7\% higher than the prior SOTA SLSDeep and converges in
less than a third of the epochs required for SLSDeep. Additionally, the
results beat Unet++, which uses the backbone of the pretrained R34 and
has a similar number of parameters to the developed model (26.1M vs
26.3M). Once more, the outcomes demonstrate the superiority of the
suggested architecture.

The effectiveness of the parallel-in-branch design and the Fusion module
is further assessed by an ablation study using various design and fusion
strategies, as shown in the below.

\begin{table}[ht]
\caption{Comparison of various fusion strategies}
\label{tab:results2}
\centering
\begin{tabular}{ccccc}
\toprule
Fusion & Jaccard & Dice & Accuracy \\
\midrule
Concat+Res & 0.778 & 0.857 & 0.939 \\
Concat+Res+CNN Spatial Attn & 0.782 & 0.861 & 0.941 \\
Concat+Res+TFM Channel Attn & 0.787 & 0.865 & 0.942 \\
Concat+Res+Dot Product & \textbf{0.795} & \textbf{0.872} & \textbf{0.944} \\
\bottomrule
\end{tabular}
\end{table}

The above table summarises the experiments performed on ISIC2017 to
support the design decision made for the Fusion module. From these
experiments, we can see that each additional component exhibits a
distinct benefit.

\subsection{Result Analysis}\label{result-analysis}
\begin{figure}
\begin{center}
\includegraphics[width=5.84769in,height=4.15151in]{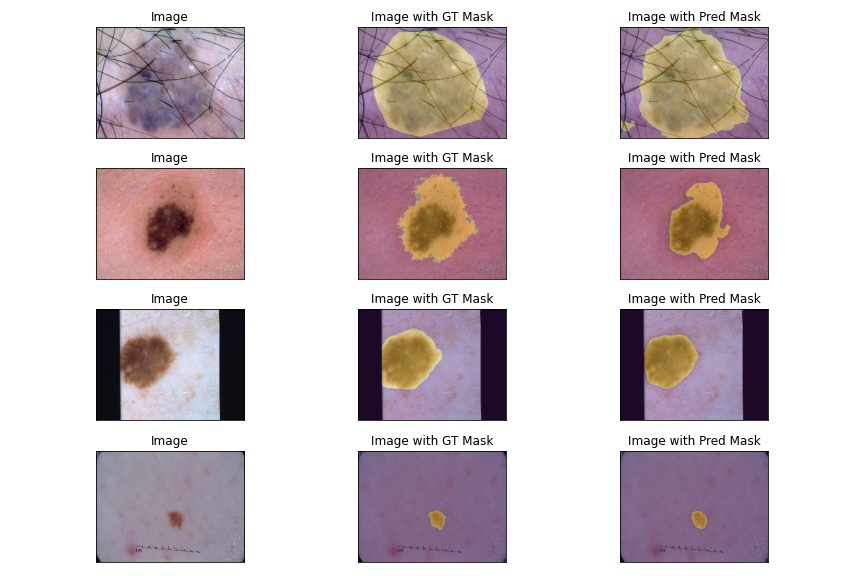}
\caption{Results visualization on 4 random Images}
\label{fig:fig4}
\end{center}
\end{figure}

We evaluate the developed model on the test images from the ISIC2017
dataset and demonstrate the effectiveness below. Visualisation results
of the model on four random images are shown in Figure \ref{fig:fig4}.

\begin{figure}
\begin{center}
\includegraphics[width=5.84769in,height=4.15151in]{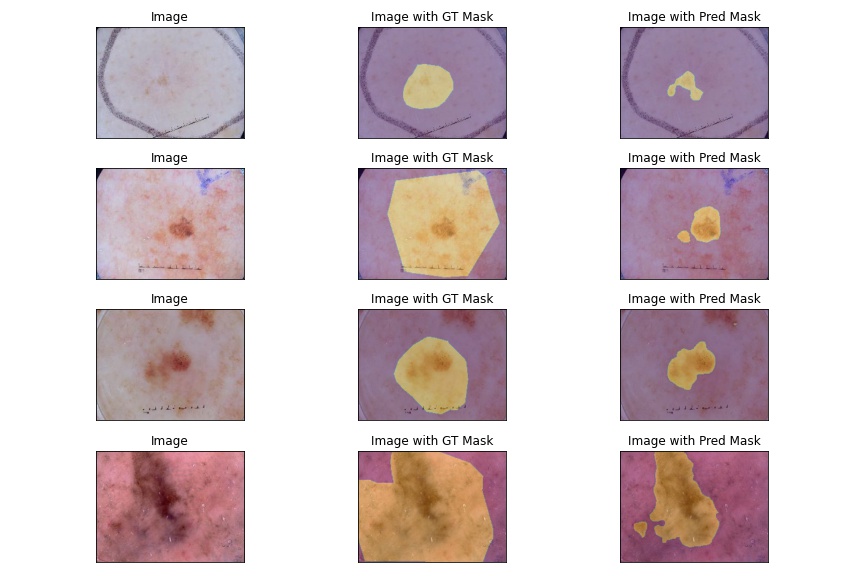}
\caption{Results visualization on 4 selected images with failures}
\label{fig:fig5}
\end{center}
\end{figure}

Visualisation results of the model where the model fails to predict a
suitable segmentation mask are also shown in Figure \ref{fig:fig5}.

From the above figures, it can be concluded that the model cannot
predict correct segmentation masks in cases where the \textbf{lesion
boundary is not strictly defined} or where there is little illumination
difference between the background and lesion area.

\section{Conclusions}\label{conclusions}

This paper presented a novel method for the segmentation of skin
lesions based on architecture having CNN and Transformer branches in
parallel, followed by a fusion module which fuses features from both
branches to make the predictions. Even though segmentation of skin
lesions is not an easy task, given the variations in the occurrence of
skin lesions as discussed in Chapter 1.1 and shown in Table \ref{tab:tab1}, it is
evident that the method developed in this research can segment the
lesions with pixel-wise accuracy of more than 94\% and that too more
efficiently than the previous state of the art methods.

This study utilised the ISIC2017 dataset. Other studies using the same
dataset, as seen in Table \ref{tab:results}, can be seen to be using very deep models
that require lots of computation resources or CNN-based sequential
architectures that require more training and cannot capture global
details. Since the research conducted in this study focused specifically
on finding a computationally and statistically efficient way for the
segmentation of skin lesions, this research can be considered to have a
positive outcome. The findings using the proposed methodology
demonstrate promising and encouraging results. Therefore, we believe
this approach will enhance the quality of further skin lesion
segmentation investigations.

Furthermore, we believe this work provides new insights into the
potential use of CNN and Transformer-based parallel architectures as an
automated feature extraction tool for supervised medical image
segmentation systems.

In this research, we have developed a novel architecture to combine
Transformers and CNNs with a novel fusion module for skin-lesion
segmentation. The model developed in this research can compete with the
state-of-the-art methods and achieve comparable performance on the
skin-lesion segmentation task while being very efficient computationally
and using much smaller parameters.

Despite the encouraging results, there are still many areas for
improvement to enhance further the results and model efficiency,
generalisation of the model and model interpretability. Those are
discussed below.

\bibliographystyle{unsrt}  
\bibliography{references_mendeley}

\end{document}